\documentclass[12pt,a4paper]{article}
\usepackage[margin=1in]{geometry}
\usepackage[T1]{fontenc}
\usepackage{amssymb}
\usepackage{amsmath}
\usepackage{xcolor}
\usepackage{cite}
\usepackage{enumitem}
\usepackage{titlesec}
\usepackage[colorlinks,allcolors=blue]{hyperref}

\titleformat{\section}
{\normalfont\fontsize{14}{14}\bfseries}{\thesection}{1em}{}

\let\OLDthebibliography\thebibliography
\renewcommand\thebibliography[1]{
  \OLDthebibliography{#1}
  \setlength{\parskip}{0pt}
  \setlength{\itemsep}{2pt plus 0.3ex}
}

\newcommand{\be}{\begin{equation}}
\newcommand{\ee}{\end{equation}}
\newcommand{\nn}{\nonumber}

\newcommand{\ad}{\text{ad}}

\newcommand{\cO}{{\cal O}}

\newcommand{\dd}{\text{d}}
\newcommand{\demi}{\frac{1}{2}}
\newcommand{\eg}{{\it e.g.}\ }
\newcommand{\ie}{{\it i.e.}\ }

\newcommand{\vir}{\mathfrak{vir}}

\begin{document}

\hrule
\begin{center}
\Large{\bfseries{{Hirota-Satsuma Dynamics as a\\
Non-Relativistic Limit of KdV Equations}}}
\end{center}
\hrule

~\\
\begin{center}
\large{Blagoje Oblak}\\
~\\
\small\it
{\tt{boblak@lpthe.jussieu.fr}}\\
Laboratoire de Physique Th\'eorique et Hautes Energies,\\
Sorbonne Universit\'e and CNRS UMR 7589, F-75005 Paris, France.
\end{center}

\vspace{-.4cm}
~\\

\begin{center}
\begin{minipage}{.95\textwidth}
\begin{center}{\bfseries{Abstract}}\end{center}
We consider a system of two coupled KdV equations (one for left-movers, the other for right-movers) and investigate its ultra-relativistic and non-relativistic limits in the sense of BMS$_3$/GCA$_2$ symmetry. We show that there is no local ultra-relativistic limit of the system with positive energy, regardless of the coupling constants in the original relativistic Hamiltonian. By contrast, local non-relativistic limits with positive energy exist, provided there is a non-zero coupling between left- and right-movers. In these limits, the wave equations reduce to Hirota-Satsuma dynamics (of type {\sc iv}) and become integrable. This is thus a situation where input from high-energy physics contributes to nonlinear science --- in this case, uncovering the limiting relation between integrable structures of KdV and Hirota-Satsuma.
\end{minipage}
\end{center}

\vspace{-.3cm}
~\\

\section{Introduction}

The Korteweg-de Vries (KdV) equation \cite{Korteweg} is a notorious nonlinear field equation that describes the slow time evolution of chiral (purely left- or right-moving) waves on a one-dimensional interface, or more general medium. (See \eg \cite{Hazewinkel} for a review of the various applications of KdV.) It is typically written in terms of a `comoving' coordinate $x^{\pm}\equiv x\pm Vt$, where $x$ is a static laboratory coordinate, $t$ is a slow time variable, and $V$ is the leading wave velocity --- in shallow water dynamics, for instance, $V$ would be a remnant of the velocity $\sqrt{gh}$ of surface gravity waves. All three quantities $x,t,V$ are dimensionless, and the KdV approximation normally holds in the limit $V\gg1$.\\

From the point of view of $(1+1)$-dimensional field theories, $V$ is akin to the speed of light in the system, and the comoving coordinates $x^{\pm}$ are really light-cone coordinates. It is then natural to wonder if there exist well-defined non-relativistic and ultra-relativistic limits of KdV, respectively corresponding to $V\to\infty$ and $V\to0$. The purpose of this note is to address this question in the case of two coupled KdV equations --- one of left-movers, the other for right-movers.\\

This investigation is motivated, firstly, by the fact that the condition $V\gg1$ is often part of the approximations that led to the KdV equation in the first place \cite{Ockendon}. It may therefore be of interest to find a non-relativistic limit ($V\to\infty$) of KdV. Secondly, a series of recent works in $2+1$ gravity \cite{Perez:2016vqo} have uncovered fall-off conditions on the metric that yield KdV equations on the space-time boundary. These fall-offs typically describe gravitational systems whose metric becomes that of Anti-de Sitter (AdS) space at infinity, with a negative cosmological constant $\Lambda\propto-V^2$. In the ultra-relativistic limit $V\to0$, the cosmological constant vanishes and space-time becomes Minkowskian, so one expects the boundary wave equation to become of Hirota-Satsuma form \cite{Hirota}, consistently with the asymptotic Bondi-Metzner-Sachs  (BMS$_3$) symmetry \cite{Bondi:1962px,Ashtekar:1996cd,Fuentealba:2017omf}.\\

We will find below that the non-relativistic and ultra-relativistic limits of KdV are only sensible if one accounts for both left- and right-movers, generally with a non-zero coupling between the two. Furthermore, we will show that the ultra-relativistic limit is always non-local, regardless of the initial, relativistic, couplings (unless one allows energy to be unbounded from below, in which case there is no problem with locality). By contrast, a local non-relativistic limit will indeed exist and yield Hirota-Satsuma dynamics of type {\sc iv} (in the sense of \cite{Sakovich}), but it will crucially require the presence of an order-one coupling between left- and right-movers. Such a coupling does not occur in the examples of KdV equations in Nature known to the author, so the limit does not appear to apply to the standard setups exhibiting KdV behaviour (unless one deliberately builds a coupled KdV system, but this is another matter).\\

Our arguments will rely on two key technical tools: the first is the formulation of KdV as a Lie-Poisson equation for the Virasoro group \cite{Khesin2003,Khesin}, and the analogous formulation of Hirota-Satsuma dynamics of type {\sc iv} as a Lie-Poisson equation for BMS$_3$, extending the type {\sc ix} setup of \cite{Fuentealba:2017omf}. The second is the limiting construction that relates Virasoro symmetry to either the ultra-relativistic group BMS$_3$ \cite{Ashtekar:1996cd,Barnich:2006av}, or to its non-relativistic twin, the Galilean Conformal Algebra in two dimensions (GCA$_2$) \cite{Bagchi:2009my}. As is probably well known to readers acquainted with these structures, BMS$_3$ and GCA$_2$ are isomorphic, so the distinction between them may seem superfluous. However, as we will show, the key difference between the definitions of ultra- and non-relativistic limits will eventually entail dramatic qualitative differences in the corresponding Lie-Poisson wave equations.\\

Note that this paper is by no means self-contained: it is assumed that the reader is familiar with all the relevant technical tools. Accordingly, we refer \eg to \cite{Khesin} for a pedagogical introduction to Lie-Poisson (or Euler-Arnold) equations, and to \cite{Oblak:2016eij} for a presentation of both the Virasoro and the BMS$_3$ groups, as well as the limits (\.In\"on\"u-Wigner contractions) relating them.\\

This work is structured as follows. In section \ref{sec2}, we set the stage by introducing the coupled KdV model to be used throughout and briefly recall how it emerges as a Lie-Poisson equation associated with the Virasoro group \cite{Khesin}. Then, in section \ref{sec3}, we derive the analogous Lie-Poisson equation for the BMS$_3$/GCA$_2$ group, resulting in a Hirota-Satsuma system \cite{Hirota} generalizing the one recently rediscovered in \cite{Fuentealba:2017omf} in the gravitational context. The sections that follow are devoted to the limits that concern us: we treat the ultra-relativistic limit first, in section \ref{sec4}, and show that it fails to yield a local Hamiltonian with bounded energy. Then we turn to the non-relativistic limit in section \ref{sec5}, and show that it goes through provided one couples the left- and right-moving KdV fields.

\section{Coupled KdV equations}
\label{sec2}

We consider two fields $T(x^+,t)$ and $\bar T(x^-,t)$, respectively belonging (at any time $t$) to the duals $\vir^*$ of two independent Virasoro algebras. Both are assumed to be $2\pi$-periodic as functions of their spatial argument $x^{\pm}$. We endow the space $\vir^*\oplus\vir^*$ with the standard Kirillov-Kostant Poisson structure. As a result, any quadratic Hamiltonian on that space leads to evolution equations of KdV type \cite{Khesin}. Specifically, we consider the Hamiltonian\footnote{Note that (\ref{ham}) is local in space even though $T$ and $\bar T$ respectively depend on two different light-cone coordinates $x^+$ and $x^-$. This suggests that the action functional is bound to be non-local, but this can be circumvented by enforcing $(V\partial_x-\partial_t)T=0$ and $(V\partial_x+\partial_t)\bar T=0$ through Lagrange multipliers.}
\be
H[T,\bar T]
=
\int_0^{2\pi}\frac{\dd x}{4\pi}\Big[A\,T(x)^2+\bar A\,\bar T(x)^2+2B\,T(x)\bar T(x)\Big]
\label{ham}
\ee
where $A,\bar A,B$ are real coefficients such that
\be
A\bar A-B^2>0\qquad\text{and}\qquad A+\bar A>0.
\label{abapo}
\ee
The matrix $\begin{pmatrix} A & B \\ B & \bar A\end{pmatrix}$ is the inverse of the inertia operator \cite{Khesin}. In practice, one can just think of $A,\bar A,B$ as coupling constants; the ultra-relativistic and non-relativistic limits will be performed by forcing these constants to scale in a certain way with the speed of light $V$.\\

The equation of motion that follows from (\ref{ham}) is obtained from the Lie-Poisson construction \cite{Khesin}, that is,
\be
(\dot T,\dot{\bar T})
=
\ad^*_{2\pi\delta H/\delta(T,\bar T)}(T,\bar T)
\label{erto}
\ee
where $\ad^*$ denotes the coadjoint representation of the Lie algebra $\vir\oplus\vir$, \ie the standard transformation law of quasi-primary fields with weight two under infinitesimal conformal transformations \cite{DiFran}. In just one Virasoro sector, one would have $\ad^*_{\epsilon}T=-\epsilon T'-2\epsilon'T+(c/12)\epsilon'''$, where $\epsilon$ is a conformal generator, $c$ is the Virasoro central charge, and the prime denotes derivatives with respect to the (spatial) argument $x$ or $x^{\pm}$. From eq.\ (\ref{erto}), we thus find
\begin{align}
\label{pdot}
\dot T(x,t)
&=
-3A\,TT'+A\frac{c}{12}T'''
-B\Big(\bar TT'+2\bar T'T-\frac{c}{12}\bar T'''\Big),\\[.2cm]
\label{pbdot}
\dot{\bar T}(x,t)
&=
-3\bar A\,\bar T\bar T'+\bar A\frac{\bar c}{12}\bar T'''
-B\Big(T\bar T'+2T'\bar T-\frac{\bar c}{12}T'''\Big),
\end{align}
where $c$ and $\bar c$ are respectively the left- and right-moving central charges. In typical situations, one considers $A,\bar A>0$ but $B=0$, which results in two decoupled, integrable KdV equations --- one for left-movers, the other for right-movers. Conversely, $B\neq0$ means there is a coupling between left- and right-movers, whereby the system (\ref{pdot})-(\ref{pbdot}) becomes non-integrable and of Hirota-Satsuma type {\sc v} \cite{Sakovich}.\\

Our goal is to take suitable (ultra- and non-relativistic) limits of the Hamiltonian (\ref{ham}) and eqs.\ (\ref{pdot})-(\ref{pbdot}), and see if there exists a scaling of $A,\bar A,B$ giving rise to finite limiting Hamiltonians that are local and bounded from below. From now on, we say that a limiting Hamiltonian is {\it consistent} if it satisfies these criteria (finiteness, locality and positivity). As we shall see, there is no consistent limiting Hamiltonian in the ultra-relativistic case, but there is one in the non-relativistic case provided $A$, $\bar A$ and $B$ are all positive, non-zero and of the same order.

\section{Hirota-Satsuma dynamics as a Lie-Poisson system}
\label{sec3}

Before we turn to the ultra- and non-relativistic limits of the coupled KdV Hamiltonian (\ref{ham}), we briefly explain how Hirota-Satsuma dynamics of type {\sc iv} emerges from the BMS$_3$ group \cite{Ashtekar:1996cd}. For a review of, and a general introduction to, the BMS$_3$ group, see \eg \cite[chap.\ 9]{Oblak:2016eij}. Its non-relativistic twin, the Galilean Conformal Group in two dimensions, was introduced in \cite{Bagchi:2009my}. (In practice one generally focusses on its Lie algebra, $\mathfrak{gca}_2$.) The two group structures are isomorphic, despite key interpretational differences that will be highlighted in due time.\\

The BMS$_3$ group is spanned by so-called `superrotations' and `supertranslations', analogously to the Euclidean and Poincar\'e groups that consist of (pseudo)rotations and translations. Abstractly, it is a semi-direct product of the form $\text{BMS}_3=\text{Diff}\,S^1\ltimes\text{Vect}\,S^1$ (up to central extensions), where $\text{Diff}\,S^1$ is the diffeomorphism group of the circle and $\text{Vect}\,S^1$ is the space of vector fields on the circle, seen here as an Abelian subgroup. The dual of the $\mathfrak{bms}_3$ algebra thus consists of pairs $(j,p)$, where both $j(x)$ and $p(x)$ are quasi-primary fields with weight two under $\text{Diff}\,S^1$. As a result, the coadjoint representation of $\mathfrak{bms}_3$ takes the form \cite{Barnich:2015uva}
\be
\ad^*_{(\epsilon,\alpha)}(j,p)
=
\Bigl(
-\epsilon j'-2\epsilon'j+\frac{c_1}{12}\epsilon'''-\alpha p'-2\alpha'p+\frac{c_2}{12}\alpha''',
-\epsilon p'-2\epsilon'p+\frac{c_2}{12}\epsilon'''
\Bigr),
\label{coad}
\ee
where $\epsilon$ and $\alpha$ are respectively infinitesimal superrotations and supertranslations, while $c_1$ and $c_2$ are the two BMS$_3$ central charges. In parity-preserving $2+1$ gravity, one has $c_1=0$ and $c_2\neq0$ \cite{Barnich:2006av}, but parity-breaking theories generally make $c_1$ non-zero as well \cite{Bagchi:2012yk}.\\

In the ultra-relativistic context, the field $p(x)$ is interpreted as an energy density while $j(x)$ is a density of angular momentum \cite{Barnich:2010eb}. Their roles are interchanged in non-relativistic systems, where $p(x)$ instead becomes a momentum density, while $j(x)$ is the energy density \cite{Bagchi:2009my}. This switch is at the root of the qualitative difference in Lie-Poisson dynamics that we will exhibit below.\\

As alluded to around eq.\ (\ref{erto}), Lie-Poisson equations are determined by the coadjoint representation of a Lie algebra, and a choice of inertia operator determining a quadratic Hamiltonian. In the present case, we choose the Hamiltonian to take the form
\be
H
=
\int_0^{2\pi}\frac{\dd x}{4\pi}\Bigl[Cp^2+2Djp+Ej^2\Bigr].
\label{cde}
\ee
where $C,D,E$ are real constants such that
\be
CE-D^2>0\qquad\text{and}\qquad C+E>0,
\label{cedepo}
\ee
ensuring that the Hamiltonian (\ref{cde}) is bounded from below. As a result, using \eqref{coad}, the Lie-Poisson wave equation $(\dot j,\dot p)=\ad^*_{2\pi\delta H/\delta(j,p)}(j,p)$ is given by
\begin{align}
\label{jidot}
\dot j
&=
-3Ejj'-3D(j'p+jp')-3Cpp'+\frac{Dc_1+Cc_2}{12}p'''+\frac{Ec_1+Dc_2}{12}j''',\\
\label{pidot}
\dot p
&=
-3Dpp'-Ejp'-2Ej'p+\frac{Dc_2}{12}p'''+\frac{Ec_2}{12}j'''.
\end{align}
This system coincides with the integrable type {\sc iv} Hirota-Satsuma equations \cite{Hirota,Sakovich}. Its special case $E=0$, also integrable, is of type {\sc ix} \cite{Sakovich} and was derived in \cite{Fuentealba:2017omf} as a boundary description of three-dimensional Chern-Simons theory with a Poincar\'e gauge group (or, equivalently, asymptotically flat $2+1$ gravity). Note that integrability would be spoiled if there were a term $jj'$ in eq.\ (\ref{pidot}), as the system would then become of type {\sc v} \cite{Sakovich} and would actually be a mere rewriting of the non-integrable coupled KdV system (\ref{pdot})-(\ref{pbdot}). The absence of $jj'$ in (\ref{pidot}) is a consequence of the lack of $j$ in the second entry of the coadjoint representation (\ref{coad}), which in turn follows from the Abelian nature of supertranslations. Thus, in this respect, commutativity ensures integrability.\\

As we now show, both the ultra-relativistic and non-relativistic limits of the coupled KdV Hamiltonian (\ref{ham}) lead to equations of motion of Hirota-Satsuma form (\ref{jidot})-(\ref{pidot}), except that energy is generally unbounded from below in the ultra-relativistic case.

\section{The ultra-relativistic limit is pathological}
\label{sec4}

It was shown in \cite{Oblak:2017ptc} that the ultra-relativistic limit of the Kirillov-Kostant Poisson structure on $\vir^*\oplus\vir^*$ is the analogous structure on the dual of $\mathfrak{bms}_3$. Thus, if the Hamiltonian \eqref{ham} has a well-defined ultra-relativistic limit, then the resulting evolution equations must be be the Lie-Poisson equation of BMS$_3$ \cite{Fuentealba:2017omf}, which coincides with the Hirota-Satsuma equation of type {\sc iv}. We now show that this limit exists, but is pathological in that one must accept to have either a non-local Hamiltonian, or one that is unbounded from below.\\

In terms of light-cone coordinates $x^{\pm}=x\pm Vt$, the ultra-relativistic limit occurs when $V\to0$. In $2+1$ gravity, $V$ is related to the cosmological constant $\Lambda$ as $V=\sqrt{-\Lambda}\equiv1/\ell$, so the ultra-relativistic limit also corresponds to the well-studied `flat limit' that relates AdS$_3$ gravity to its Minkowskian analogue \cite{Barnich:2012aw}. Following the latter reference, the ultra-relativistic limit of a CFT stress tensor is obtained by defining
\be
T(x)
=
\demi\bigg[j(x)+\frac{p(x)}{V}\bigg],
\qquad
\bar T(x)
=
\demi\bigg[-j(-x)+\frac{p(-x)}{V}\bigg]
\label{urli}
\ee
and letting $V\to0$. The fields $p(x)$ and $j(x)$, defined with these scalings, play an important role for BMS$_3$ symmetry, as they respectively generate supertranslations and superrotations \cite{Barnich:2015uva,Oblak:2016eij}. In the present context, however, one can simply think of (\ref{urli}) as a field redefinition. The redefinition is non-local, by construction, but this turns out to be unavoidable in the ultra-relativistic regime \cite{Barnich:2012aw}.\\

Plugging (\ref{urli}) into the Hamiltonian \eqref{ham}, we see that the term $BT\bar T$ contains non-local products $j(x)p(-x)$. There is no change of variables that removes this non-locality while also preserving the definition (\ref{urli}). Thus, if we require that the limiting Hamiltonian be local in terms of the $(j,p)$ fields as defined in (\ref{urli}), we are forced to set
\be
B=0.
\label{bzero}
\ee
As a result, the Hamiltonian boils down to the following local expression (all fields being evaluated at the same point $x$):
\be
H=
\int_0^{2\pi}\frac{\dd x}{16\pi}\Bigl[(A+\bar A)\frac{p^2}{V^2}+2(A-\bar A)\frac{jp}{V}+(A+\bar A)j^2\Bigr].
\nn
\ee
We now require all terms in this Hamiltonian to be finite in the limit $V\to0$. This imposes the scalings
\be
A+\bar A\sim C V^2,
\qquad
A-\bar A\sim D V
\label{aba}
\ee
in the small $V$ limit, where $C,D$ are finite ($V$-independent) constants. In the limit, the term $CV^2$ is negligible compared to $DV$, so eq.\ \eqref{aba} says that $A,\bar A=\cO(V)$ with
\be
A\sim-\bar A+\cO(V^2).
\label{nega}
\ee
This, in turn, implies that either $A$ or $\bar A$ is negative in the limit (but not both), which, given $B=0$, contradicts our basic assumption \eqref{abapo}.\\

Conclusion: there exists no consistent ultra-relativistic limit of two coupled KdV equations. The root of this negative result is the (standard \cite{Barnich:2012aw}) non-locality of the field redefinition (\ref{urli}). Note that, by relaxing one of our assumptions on consistency, one can in fact obtain a Hirota-Satsuma Hamiltonian, albeit a mildly pathological one. Indeed, if we do {\it not} require energy to be positive, then eq.\ \eqref{nega} is perfectly acceptable; the equation of motion obtained by taking the $V\to 0$ limit of eqs.\ (\ref{pdot})-(\ref{pbdot}) coincide with eqs.\ (\ref{jidot})-(\ref{pidot}) with $E=0$ and the BMS$_3$ central charges
\be
c_1=\lim_{V\to 0}(c-\bar c),
\qquad
c_2=\lim_{V\to 0}V(c+\bar c).
\label{c1c2}
\ee
In that sense, the BMS$_3$ Lie-Poisson equation found in \cite{Fuentealba:2017omf} (which has $E=0$) can be recovered by suitably scaling the KdV inertia operator, the price being that the Hamiltonian is no longer positive-definite. In fact, we now know that there exists no `flat limit', whatsoever, of KdV dynamics in AdS$_3$ \cite{Perez:2016vqo} that would yield a limiting theory with positive energy. This may have implications for boundary fluid dynamics, as studied \eg in \cite{Ciambelli:2018xat}.

\section{The non-relativistic limit is consistent}
\label{sec5}

The non-relativistic limit is similar to the ultra-relativistic one, save for a key difference in locality that eventually makes the non-relativistic limit much better behaved. Namely, one defines the non-relativistic limit of the CFT stress tensor $(T,\bar T)$ by letting \cite{Bagchi:2009my}
\be
T(x)
=
\demi\Bigl(j(x)+V\,p(x)\Bigr),
\qquad
\bar T(x)
=
\demi\Bigl(j(x)-V\,p(x)\Bigr)
\label{nrli}
\ee
and taking the limit $V\to\infty$. Note the key differences between this equation and its ultra-relativistic analogue \eqref{urli}:
\begin{enumerate}[itemsep=0cm]
\item[(i)] there is no inversion $x\mapsto-x$, so the redefinition is manifestly local;
\item[(ii)] the minus sign in the second equation is in front of $p$ rather than $j$, so $j(x)$ is now an energy density while $p(x)$ is an (angular) momentum density;
\item[(iii)] the speed of light $V$ multiplies (rather than divides) $p(x)$, but the end result is the same: the contribution of $p(x)$ to $T$ and $\bar T$ is dominant with respect to that of $j(x)$.
\end{enumerate}
Recall that, before taking the limit, the fields $T,\bar T$ are respectively functions of $x^+$ and $x^-$ only. In the limit $V\to\infty$, one therefore expects them to become functions of $t$ alone (not of $x$). If so, the theory would automatically become non-local in space. Our point of view here is that, instead, one should view $T,\bar T$ as abstract functions of some coordinate $x$ (without specification as to what that coordinate is), whereby eq.\ (\ref{nrli}) is merely a field redefinition with $2\pi$-periodic fields $j(x),p(x)$.\\

Plugging the redefinition \eqref{nrli} in the Hamiltonian \eqref{ham}, we find
\be
H
=
\int_0^{2\pi}\frac{\dd x}{16\pi}\Bigl[\bigl(A+\bar A-2B\bigr)V^2p^2+2(A-\bar A)Vjp+\bigl(A+\bar A+2B\bigr)j^2\Bigr].
\label{hlim}
\ee
As announced, in contrast to the ultra-relativistic case, there is now no issue with locality, and we are free to keep $B\neq0$. (To be compared to the previous condition \eqref{bzero}.) We now take the limit $V\to\infty$ and ask whether there exists a scaling of $A,\bar A,B$ with $V$ such that the limiting Hamiltonian \eqref{hlim} is consistent. To answer this, we define finite ($V$-independent) constants $C,D,E$ such that
\begin{align}
A&\sim E+\frac{2D}{V}+\frac{C}{V^2}+\cO(|V|^{-3}),\nn\\[.2cm]
\label{sca}
\bar A&\sim E-\frac{2D}{V}+\frac{C}{V^2}+\cO(|V|^{-3}),\\[.2cm]
B&\sim E-\frac{C}{V^2}+\cO(|V|^{-3})\nn
\end{align}
in the limit $V\to\infty$. This scaling automatically gives rise to the finite, local limiting Hirota-Satsuma Hamiltonian (\ref{cde}). The latter is bounded from below if and only if conditions (\ref{cedepo}) hold, which are the limiting analogues of eq.\ \eqref{abapo}. One readily verifies that \eqref{cedepo} is indeed satisfied provided $A,\bar A,B$ satisfy \eqref{abapo}.\footnote{Indeed, given \eqref{sca}, one has $0<A+\bar A\sim 2E$ and $0<A\bar A-B^2\sim4(CE-D^2)/V^2$. The former condition yields $E>0$; the latter gives $CE-D^2>0$, which in turn yields $C>0$, hence $C+E>0$.}\\

In conclusion, there exists a consistent non-relativistic limit of two coupled KdV equations {\it provided} their coupling $B$ is non-zero, positive, and of the same order as the self-couplings $A,\bar A$. This proviso follows from the scalings \eqref{sca}, and the resulting limiting equations of motion take the form (\ref{jidot})-(\ref{pidot}) with central charges
\be
c_1=\lim_{V\to\infty}(c+\bar c),
\qquad
c_2=\lim_{V\to\infty}\frac{c-\bar c}{V}.
\ee
(Note the difference between these relations and eqs.\ (\ref{c1c2}), mimicking the difference between ultra- and non-relativistic redefinitions of the stress tensor.) We stress again that the limiting equation of motion (\ref{pidot}) has no term $jj'$ on the right-hand side, ensuring that the system (\ref{jidot})-(\ref{pidot}) is of type {\sc iv} and integrable \cite{Sakovich}. This cancellation stems from the fact that the contribution of $j(x)$ is subleading with respect to that of $p(x)$ in the redefinition (\ref{nrli}), so that the limit $V\to\infty$ sets the coefficient of $jj'$ to zero in the equation for $\dot p$. By contrast, prior to the limit (\ie at finite $V$), there is a non-zero $jj'$ term on the right-hand side of (\ref{pidot}) and the system of equations for $(j,p)$ loses integrability \cite{Sakovich}. In this sense, the non-relativistic limit of coupled KdV equations is singular, as it turns a non-integrable system (type {\sc v} Hirota-Satsuma) into an integrable one (type {\sc iv}). This singularity is an echo of BMS$_3$/GCA$_2$ supertranslations, which indeed become commutative only {\it after} taking an \.In\"on\"u-Wigner  contraction of two Virasoro algebras \cite[sec.\ 9.4]{Oblak:2016eij}.\\

Note that order-one couplings between $T$ and $\bar T$, as in eqs.\ (\ref{sca}), definitely do not occur in standard fluid dynamics \cite{Ockendon}; nor do they appear in any instance of KdV in physics known to the author. It may therefore be interesting to uncover systems exhibiting this kind of behaviour. In the context of $2+1$ gravity, this limit would yield a non-relativistic version of the KdV boundary dynamics found in \cite{Perez:2016vqo}. It would be interesting to see what `gravity-like' theory has couplings satisfying the scalings (\ref{sca}): such a theory would have a well-defined non-relativistic limit, suggesting applications in non-relativistic AdS/CFT \cite{Bagchi:2009my,Harmark:2018cdl}.

\section*{Acknowledgements}

I am grateful to G.\ Kozyreff for support and collaboration on related projects, and to H.\ Gonz\'alez and M.\ Pino for discussions that led to the writing of this note. Many thanks also to the anonymous referee for insightful comments on the first version of this paper. This work is supported by the ANR grant {\it TopO}, number ANR-17-CE30-0013-01.

\addcontentsline{toc}{section}{References}

\end{document}